\documentclass{article}
\usepackage[hyphens]{url}
\usepackage[utf8]{inputenc}
\usepackage{graphicx}
\usepackage{authblk}
\usepackage[english]{babel}
\usepackage{setspace}
\usepackage{fancyhdr}
\usepackage{array}
\usepackage[margin=1in]{geometry}
\usepackage{float}
\usepackage{amsmath}
\usepackage{amsthm}
\usepackage{amssymb}
\usepackage{bbm}
\usepackage{bm}
\usepackage{scalerel,stackengine}
\usepackage[final]{pdfpages}
\usepackage{longtable}
\usepackage{chngcntr}
\usepackage{placeins}
\usepackage{microtype}
\usepackage{tikz}
\usepackage{makecell}
\usepackage{hyperref}
\usepackage{subcaption}
\usepackage{rotating}
\usepackage{soul}
\usepackage{tabularx}
\usepackage{wasysym}

\theoremstyle{definition}

\hypersetup{
    colorlinks = true,
    citecolor = {blue},
    urlcolor = {blue},
    menucolor = {blue},
    linkcolor = {black}
}

\usetikzlibrary{shapes, decorations, positioning, calc}

\newcommand{\probP}{\text{I\kern-0.15em P}}

\title{Finite-sample correction for the covariate-adjusted log-rank test}

\author[1]{Pavla Krotka}
\author[2]{Dominic Magirr}
\affil[1]{Department of Statistics and Operations Research and Institute for Research and Innovation in Health (IRIS), Universitat Politècnica de Catalunya - BarcelonaTech (UPC), Barcelona, Spain}
\affil[2]{Novartis Pharma AG, Basel, Switzerland}
\date{}         
\begin{document}

\maketitle

\begin{abstract}
    The covariate-adjusted log-rank test is a novel method for covariate adjustment in randomized trials with time-to-event endpoints, offering guaranteed efficiency gains compared to the standard log-rank test. However, it has been noted that, in small samples, this method may lead to type I error rate inflation. This issue is particularly pronounced in trials with imbalanced allocation and settings where the number of adjustment covariates is large relative to the sample size. We propose a finite-sample correction for the denominator of the covariate-adjusted log-rank test statistic that accounts for the loss of the residual degrees of freedom as well as the uncertainty in the unknown regression coefficients. In simulations, we show that applying this correction leads to a substantial reduction in the type I error rate inflation across multiple scenarios.
\end{abstract}

\section{Introduction}

The covariate-adjusted log-rank test proposed by Ye et al. \cite{ye2024covariate} is a powerful method for covariate adjustment in survival trials targeting the marginal hazard ratio. This approach offers guaranteed efficiency gains compared to the standard log-rank test and is asymptotically valid even under covariate-adaptive randomization. Moreover, a user-friendly software implementation of the methodology is provided in the \texttt{RobinCar2} R package \cite{bannick2026robincar}, facilitating its employment in clinical trial practice.

However, it has recently been noted that the covariate-adjusted log-rank test may lead to type I error rate inflation in finite samples \cite{bannick2026covariate, backenroth2026practical}. This is particularly pronounced in imbalanced trials and settings where the number of adjustment covariates is large relative to the sample size. Aiming to eliminate this inflation, we propose a correction for the estimator of the variance of the numerator in the covariate-adjusted log-rank test statistic that properly accounts for the reduction in the residual degrees of freedom and the variance inflation factor due to estimation of unknown regression parameters -- two sources of variability recently highlighted by Senn et al. \cite{senn2026stratification, senn2025covariate}. We demonstrate via simulations that applying the correction substantially reduces the type I error rate compared to the original covariate-adjusted log-rank test across multiple simulation scenarios with varying sample sizes and allocation ratios.

\section{Finite-sample correction}

Consider a randomized clinical trial evaluating the efficacy of an experimental arm against a control arm, with group sizes $n_1$ and $n_0$, respectively. Denote the treatment assignment for patient $i$ ($i = 1, \ldots, n=n_1 + n_0$) by $A_i \in \{0,1\}$. Suppose that a $k$-dimensional vector of covariates, denoted $X_i$, has been observed for each patient. In a standard time-to-event analysis, the log-rank test statistic is
\begin{equation*}
    T_L = \hat{U}_L  / \sqrt{\widehat{\operatorname{Var}}}(\hat{U}_L)
\end{equation*}
where the numerator is a sum over event times $t_j$ of observed $O_{1,j}$ minus expected $ E_{1,j}$ events on treatment arm $A=1$ under the null assumption of equal survival curves on both arms,
\begin{equation*}
\hat{U}_L = \sum_{t_j} O_{1,j} - E_{1,j}.
\end{equation*}
Ye et al. \cite{ye2024covariate} construct a covariate-adjusted log-rank test statistic that makes use of the covariate information, $X_i~(i=1,\ldots,n)$,  
\begin{equation*}\label{eq_tcl}
    T_{CL} = \hat{U}_{CL}  / \sqrt{\widehat{\operatorname{Var}}}(\hat{U}_{CL})
\end{equation*}
Asymptotically, $T_{CL}$ controls the type I error rate and is guaranteed to be no less efficient than $T_L$. However, it can be shown that, in finite samples,
\begin{equation*}
    \mathbb{E}\left\lbrace\widehat{\operatorname{Var}}(\hat{U}_{CL})\right\rbrace < \operatorname{Var}(\hat{U}_{CL}) 
\end{equation*}
leading to type I error inflation.
The aim, therefore, is to find a finite-sample correction term, $\Gamma$, such that
\begin{equation}\label{eq_aim}
    \mathbb{E}\left\lbrace\Gamma\cdot \widehat{\operatorname{Var}}(\hat{U}_{CL})\right\rbrace \approx \operatorname{Var}(\hat{U}_{CL}). 
\end{equation}

\subsection{Main result}

For multivariate Gaussian $X_i$, the following correction term achieves (\ref{eq_aim}),

\begin{equation*}
\Gamma (n_1, n_0, k) = \left\lbrace 
\frac{n}{n - \left( \frac{n_0}{n_1} + \frac{n_1}{n_0} -1\right)k - \left( \frac{n_0}{n_1} + \frac{n_1}{n_0} \right) } \right\rbrace \left\lbrace 1 + k \left( \frac{(n_0/n)^2}{n_1 - k - 2} + \frac{(n_1/n)^2}{n_0 - k - 2} \right) \right\rbrace.
\end{equation*}

\subsection{Derivation}

To derive $\Gamma(n_1, n_0, k)$, our strategy is to consider a linear model with treatment-covariate interactions, and establish by how much a plug-in estimator for the variance of the least-squares estimator of average treatment effect  underestimates the true variance (assuming multivariate normal covariates). Since the same structure is used in the regression adjustment in Ye et al. \cite{ye2024covariate}, and the same plug-in type variance estimator is used, it follows that the finite-sample correction derived in the linear model case can be transferred to the covariate-adjusted log-rank test. 

To elaborate, Ye et al. first obtain an asymptotic linear representation of the log-rank test statistic in terms of patient-level derived outcomes. That is, the numerator of the log-rank statistic, $\hat{U}_L$ can be expressed as a between-arm contrast of the means of the derived outcomes.  To derive the numerator of the adjusted log-rank statistic, $\hat{U}_{CL}$, these derived outcomes are regressed on the baseline covariates separately within each arm, and the resulting arm-specific regressions are evaluated at their arm-specific covariate means before taking the same contrast as the original procedure. In other words, $\hat{U}_L$ is analogous to a difference in sample means when fitting a simple linear model with treatment term only, while $\hat{U}_{CL}$ is analogous to the estimated average treatment effect from a linear model containing all covariates as main effects and as treatment-covariate interaction effects.

The linear model with treatment-covariate interactions is given by:
\begin{equation*}
 \mathbb E (Y_i \mid A_i, X_i) = \nu_1 A_i + \nu_0 (1-A_i) + \beta_1^\top X_i A_i + \beta_0^\top X_i (1 - A_i) 
\end{equation*}
where $\nu_1$ and $\nu_0$ are the intercepts for the treatment and control arm, respectively, and $\beta_1$ and $\beta_0$ are the covariate coefficients estimated separately for each arm. Note that this corresponds to fitting a separate linear regression model in each arm.
Denoting the overall covariate mean by $\bar{X}$, the OLS estimator of the average treatment effect is given by:
\begin{equation*}
    \hat \theta = \hat{\nu}_1 - \hat{\nu}_0 + (\hat{\beta}_1 - \hat{\beta}_0)^\top \bar{X}
\end{equation*}
If we assume homogeneous residual variance $\sigma^2$, we have, 
\begin{equation*}
    \operatorname{Var} (\hat \theta \mid \mathbf{X}) = \sigma^2 c^\top \!(\mathbf{X}^\top \mathbf{X})^{-1}c = \sigma^2 \left( \frac{1}{n_1} + \frac{1}{n_0} \right) \cdot \mathrm{VIF},
\end{equation*}
where $\mathbf{X}$ is the design matrix,  $\mathrm{VIF} = c^\top \!(\mathbf{X}^\top \mathbf{X})^{-1}c / (1/n_1 + 1/n_0)$ is the variance inflation factor \cite{senn2026stratification}, and $c=(1,-1,\bar X^\top,-\bar X^\top)$ is a contrast vector.
If we focus on null scenarios where $\beta_1 = \beta_0$, then, by the law of total variance,
\begin{equation*}
\begin{split}
   \operatorname{Var}(\hat\theta)&=\mathbb{E}(\operatorname{Var}(\hat\theta\mid \mathbf{X})) +\operatorname{Var}(\mathbb{E}(\hat\theta\mid \mathbf{X}))  \\
   &=\mathbb{E}(\operatorname{Var}(\hat\theta\mid \mathbf{X})) + (\beta_1-\beta_0)^\top \operatorname{Var}(\bar X) (\beta_1-\beta_0) \\
   &=\sigma^2\mathbb E\left[\left( \frac{1}{n_1} + \frac{1}{n_0} \right) \cdot \mathrm{VIF}\right] 
   \end{split}
\end{equation*}
In the following, we will condition on the observed sample sizes $n_0$ and $n_1$ so we can consider them fixed. Then, 
\begin{equation}\label{eq_var_theta_hat}
     \operatorname{Var}(\hat\theta) = \sigma^2\mathbb E ( \mathrm{VIF} ) \left( \frac{1}{n_1} + \frac{1}{n_0} \right)
\end{equation}
For linear models, Ye et al. proposed the following plug-in variance estimator \cite{ye2023toward}:\begin{equation}\label{eq_var_robin_lm}
     \widehat{\operatorname{Var}}(\hat\theta) = \frac{\hat{\sigma}_1^2}{n_1} + \frac{\hat{\sigma}_0^2}{n_0} + (\hat{\beta}_1-\hat{\beta_0})^\top \frac{\widehat{\operatorname{Var}} (X)}{n}(\hat{\beta_1}-\hat{\beta_0}).
\end{equation}
We shall consider $\mathbb E( \widehat{\operatorname{Var}}(\hat\theta))$ under a multivariate normal model and compare with (\ref{eq_var_theta_hat}). If $\hat{\sigma}_j^2$ is a maximum likelihood estimator of $\sigma^2$, then the expectation of the first two terms in (\ref{eq_var_robin_lm}) is 
\begin{equation*}
    \frac{\sigma^2}{n_j}\left(  \frac{n_j - k - 1}{n_j} \right).
\end{equation*}
In the Appendix, we show that the expected value of the third term in (\ref{eq_var_robin_lm}) is approximately equal to 
\begin{equation*}
    \frac{\sigma^2k}{n}\left(  \frac{1}{n_1} +  \frac{1}{n_0}\right).
\end{equation*}
After rearrangement, we obtain 
\begin{equation*}
    \mathbb E( \widehat{\operatorname{Var}}(\hat\theta)) \approx \sigma^2 \left\lbrace \frac{n - \left( \frac{n_0}{n_1} + \frac{n_1}{n_0} -1\right)k - \left( \frac{n_0}{n_1} + \frac{n_1}{n_0} \right) }{n} \right\rbrace \left(   \frac{1}{n_1} +  \frac{1}{n_0}\right)
\end{equation*}
so that
\begin{equation*}
    \mathbb E\left( C_{df}(n_0,n_1,k) \cdot \mathbb E (\mathrm{VIF}) \cdot \widehat{\operatorname{Var}}(\hat\theta) \right) \approx\sigma^2 \mathbb E (\mathrm{VIF}) \left( \frac{1}{n_1} + \frac{1}{n_0} \right) = \operatorname{Var}(\hat\theta),
\end{equation*}
where 
\begin{equation*}
    C_{df}(n_0,n_1,k) = \frac{n}{n - \left( \frac{n_0}{n_1} + \frac{n_1}{n_0} -1\right)k - \left( \frac{n_0}{n_1} + \frac{n_1}{n_0} \right) }
\end{equation*}
is the correction for the loss of residual degrees of freedom. Finally, we show in the Appendix that, under a multivariate normal model,
\begin{equation*}
    \mathbb E \left(  \mathrm{VIF} \right)=\left\lbrace 1 + k \left( \frac{(n_0/n)^2}{n_1 - k - 2} + \frac{(n_1/n)^2}{n_0 - k - 2} \right) \right\rbrace .
\end{equation*}



\section{Simulations}\label{sec:simulations}

We perform a simulation study to evaluate the performance of the covariate-adjusted log-rank test and the proposed correction for finite samples. We compare $\mathbb{E}\left\lbrace  \widehat{\operatorname{Var}}(\hat{U}_{CL})\right\rbrace$  and  $\mathbb{E}\left\lbrace\Gamma\cdot \widehat{\operatorname{Var}}(\hat{U}_{CL})\right\rbrace$ with the empirical variance of $\hat{U}_{CL}$. And we assess the type I error rate across multiple scenarios, varying the sample sizes, allocation ratios, and the number of adjustment covariates.

\subsection{Data generation and considered scenarios}

Covariates were simulated from a multivariate normal distribution with mean zero and identity variance matrix. Survival times were generated independently from the covariates. The survival times on both treatment arms were generated from an exponential distribution $T \sim \exp(\gamma)$, where $\gamma = \log(2) / t^{\text{med}}$ and $t^{\text{med}}$ is the median survival time. The recruitment period was set to 6 months, and administrative censoring was included, with the end of study set to 18 months. Patients without an observed event at the end of the study were censored.

We focus on varying the number of adjustment covariates, as well as investigating different sample sizes and allocation ratios. In particular, we consider three different total sample sizes: small ($n=200$), moderate ($n=500$), and large ($n=4000$) trials. For each total sample size, we investigate equal allocation, as well as trials with 3:2, 2:1 and 3:1 allocation ratios. The considered simulation scenarios are summarized in Table \ref{tab:scenarios}. All scenarios were replicated 100.000 times.

\begin{table}[H]
\resizebox{\textwidth}{!}{\begin{tabular}{|c|c|c|c|c|c|c|}
\hline
                              & \textbf{$n$} & \textbf{Allocation ratio} & \textbf{$k$} & \textbf{\begin{tabular}[c]{@{}c@{}}Median \\ survival time\end{tabular}} & \textbf{\begin{tabular}[c]{@{}c@{}}Recruitment \\ time\end{tabular}} & \textbf{\begin{tabular}[c]{@{}c@{}}End \\ of study\end{tabular}}  \\ \hline
\textbf{Small sample size} & 200          & 1:1, 3:2, 2:1, 3:1        & 1,3,5,10         & 12                                                                       & 6                                                                    & 18                                                                                \\ \hline
\textbf{Moderate sample size}    & 500          & 1:1, 3:2, 2:1, 3:1        & 1,3,5,10         & 12                                                                       & 6                                                                    & 18                                                                              \\ \hline
\textbf{Large sample size}    & 4000         & 1:1, 3:2, 2:1, 3:1        & 1,3,5,10         & 60                                                                       & 6                                                                    & 18                                                                                \\ \hline
\end{tabular}}
\caption{Scenarios considered in the simulation study.}
\label{tab:scenarios}
\end{table}

\subsection{Results}

\begin{figure}[H]
    \centering
    \includegraphics[width=\linewidth]{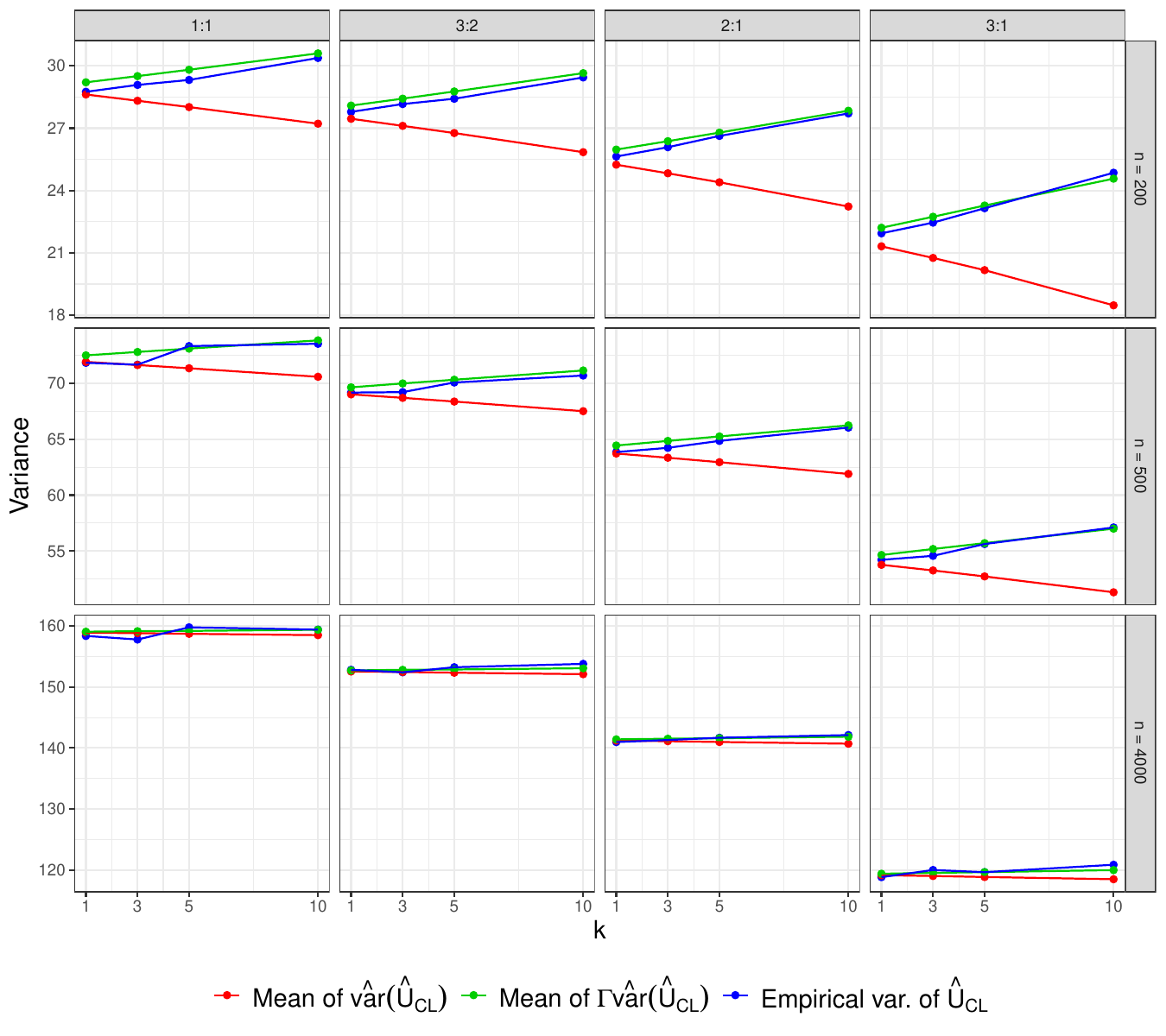}
    \caption{Empirical estimates of $\mathbb{E}\left\lbrace  \widehat{\operatorname{Var}}(\hat{U}_{CL})\right\rbrace$  and  $\mathbb{E}\left\lbrace\Gamma\cdot \widehat{\operatorname{Var}}(\hat{U}_{CL})\right\rbrace$, versus the empirical variance of $\hat{U}_{CL}$ for varying number of adjustment covariates $k$. Results for different allocation ratios are shown in the columns, while different total sample sizes are presented in the rows. Based on 100.000 simulation runs.}
        \label{fig:results_var}
\end{figure}

Figure \ref{fig:results_var} shows empirical estimates of $\mathbb{E}\left\lbrace  \widehat{\operatorname{Var}}(\hat{U}_{CL})\right\rbrace$  and  $\mathbb{E}\left\lbrace\Gamma\cdot \widehat{\operatorname{Var}}(\hat{U}_{CL})\right\rbrace$, versus the empirical variance of $\hat{U}_{CL}$. The correction term leads to much closer agreement.

\begin{figure}[H]
    \centering
    \includegraphics[width=\linewidth]{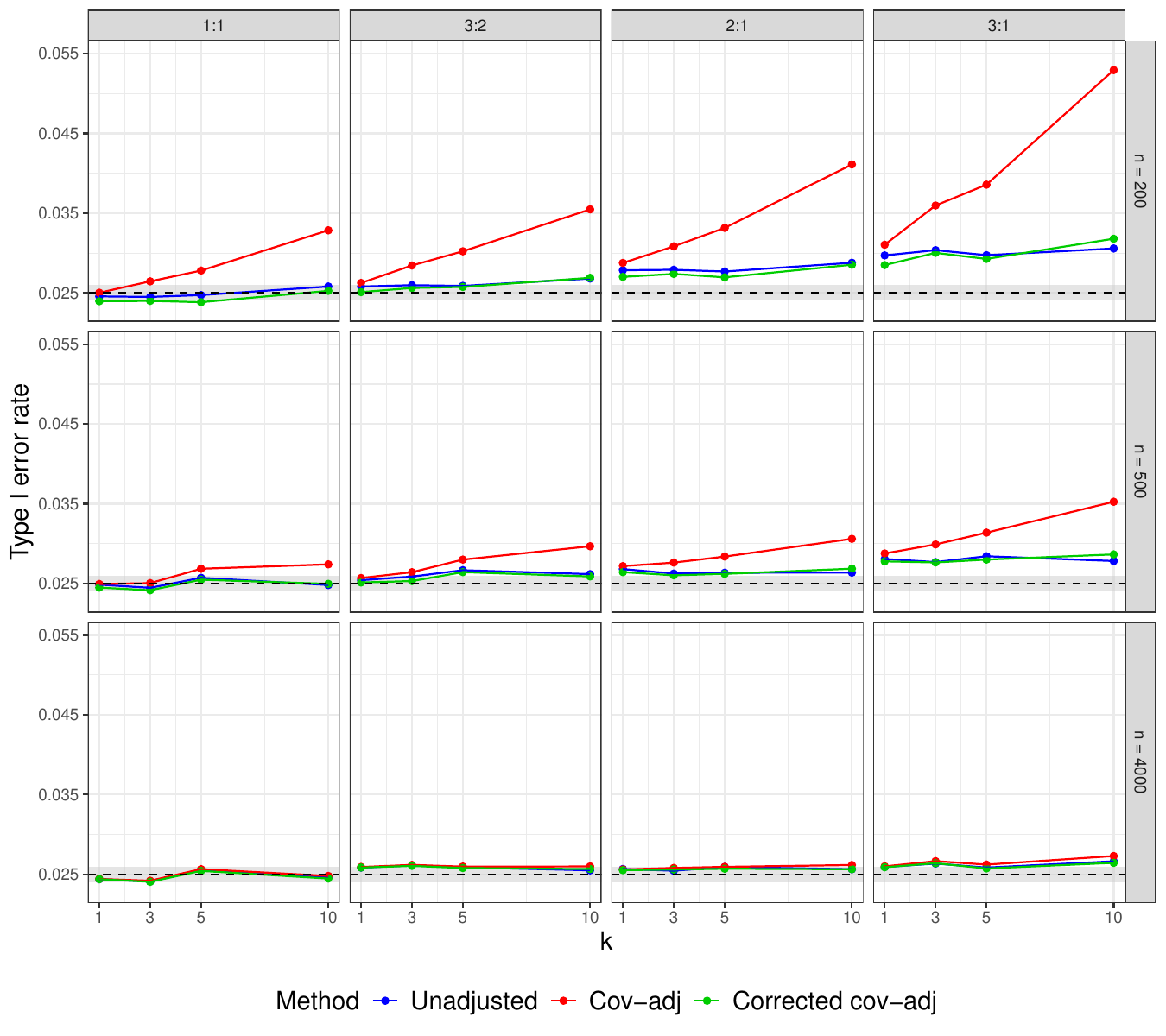}
    \caption{Type I error rate of the covariate-adjusted log-rank test, the corrected covariate-adjusted log-rank test, and the standard log-rank test for varying number of adjustment covariates $k$. Results for different allocation ratios are shown in the columns, while different total sample sizes are presented in the rows. Each plot includes a dashed reference line for the nominal significance level of 0.025, and the simulation error is shown as a gray area representing the 95\% confidence interval of the simulated type I error rate using 100.000 simulation runs.}
    \label{fig:results_t1e}
\end{figure}

Figure \ref{fig:results_t1e} shows the simulated one-sided type I error rate of the uncorrected and corrected versions of the covariate-adjusted log-rank test, applied at a nominal type I error rate of 0.025. For comparison, we also include results for the unadjusted log-rank test.  As the number of adjustment covariates increases, the uncorrected version of the covariate-adjusted log-rank test leads to a non-ignorable type I error rate inflation in cases with small sample sizes ($n=200$) and moderate sample sizes ($n=500$) with an imbalanced allocation ratio. This inflation is substantially reduced when applying the proposed finite-sample correction, although the type I error rate is not fully maintained at the nominal level in settings with highly imbalanced allocation. However, even in these cases, the performance of the corrected log-rank test is comparable to the unadjusted log-rank test.
In scenarios with large sample sizes ($n=4000$), the original method maintains the type I error rate, and the correction only has a negligible effect.

\section{Extensions}

\begin{figure}[H]
    \centering
    \includegraphics[width=\linewidth]{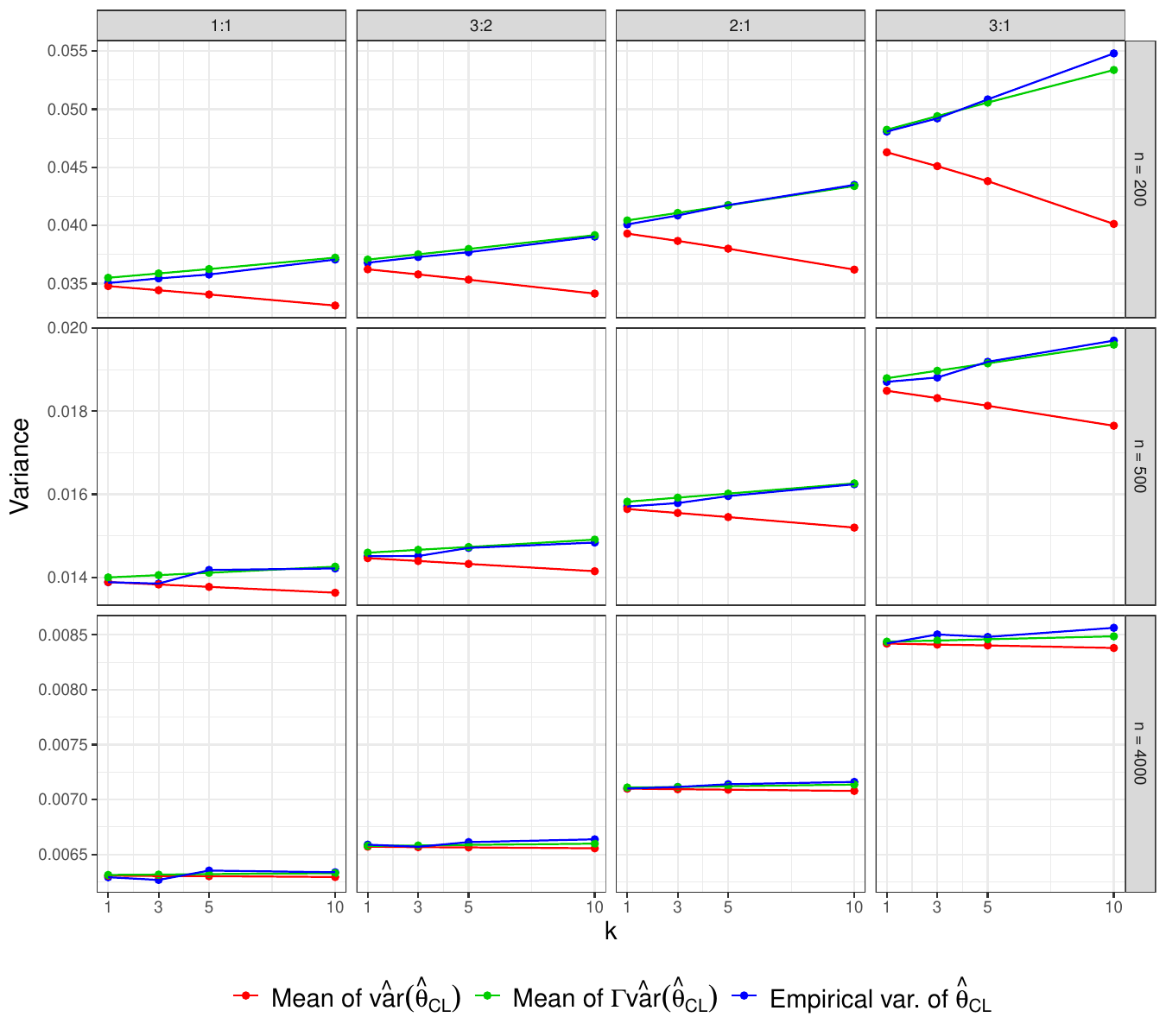}
    \caption{Empirical estimates of $\mathbb{E}\left\lbrace  \widehat{\operatorname{Var}}(\hat{\theta}_{CL})\right\rbrace$  and  $\mathbb{E}\left\lbrace\Gamma\cdot \widehat{\operatorname{Var}}(\hat{\theta}_{CL})\right\rbrace$, versus the empirical variance of $\hat{\theta}_{CL}$ for varying number of adjustment covariates $k$. Results for different allocation ratios are shown in the columns, while different total sample sizes are presented in the rows. Based on 100.000 simulation runs.}
        \label{fig:results_var_lhr}
\end{figure}

As well as the covariate-adjusted log-rank test, Ye et al. (\cite{ye2024covariate}) also propose a covariate-adjusted estimator $\hat\theta_{CL}$ for the marginal log-hazard ratio, together with an associated variance estimator  $\widehat{\operatorname{Var}}(\hat{\theta}_{CL})$. This variance estimator is proportional to $\widehat{\operatorname{Var}}\left\lbrace \hat{U}_{CL}(\hat\theta_{CL}))\right\rbrace$, where $\widehat{\operatorname{Var}}\left\lbrace \hat{U}_{CL}(\hat\theta_{CL})\right\rbrace$ is constructed in the same way as $\widehat{\operatorname{Var}}(\hat{U}_{CL})$ but under an assumption that the marginal log-hazard ratio is $\hat\theta_{CL}$  instead of zero. Therefore, the same finite-sample correction $\Gamma$ can be applied. In Figure \ref{fig:results_var_lhr} we show the agreement between the empirical variance of $\hat{\theta}_{CL}$ and the empirical mean of $\widehat{\operatorname{Var}}(\hat{\theta}_{CL})$ with and without correction $\Gamma$ using the same scenarios as in Section \ref{sec:simulations}.

\begin{figure}[H]
    \centering
    \includegraphics[width=\linewidth]{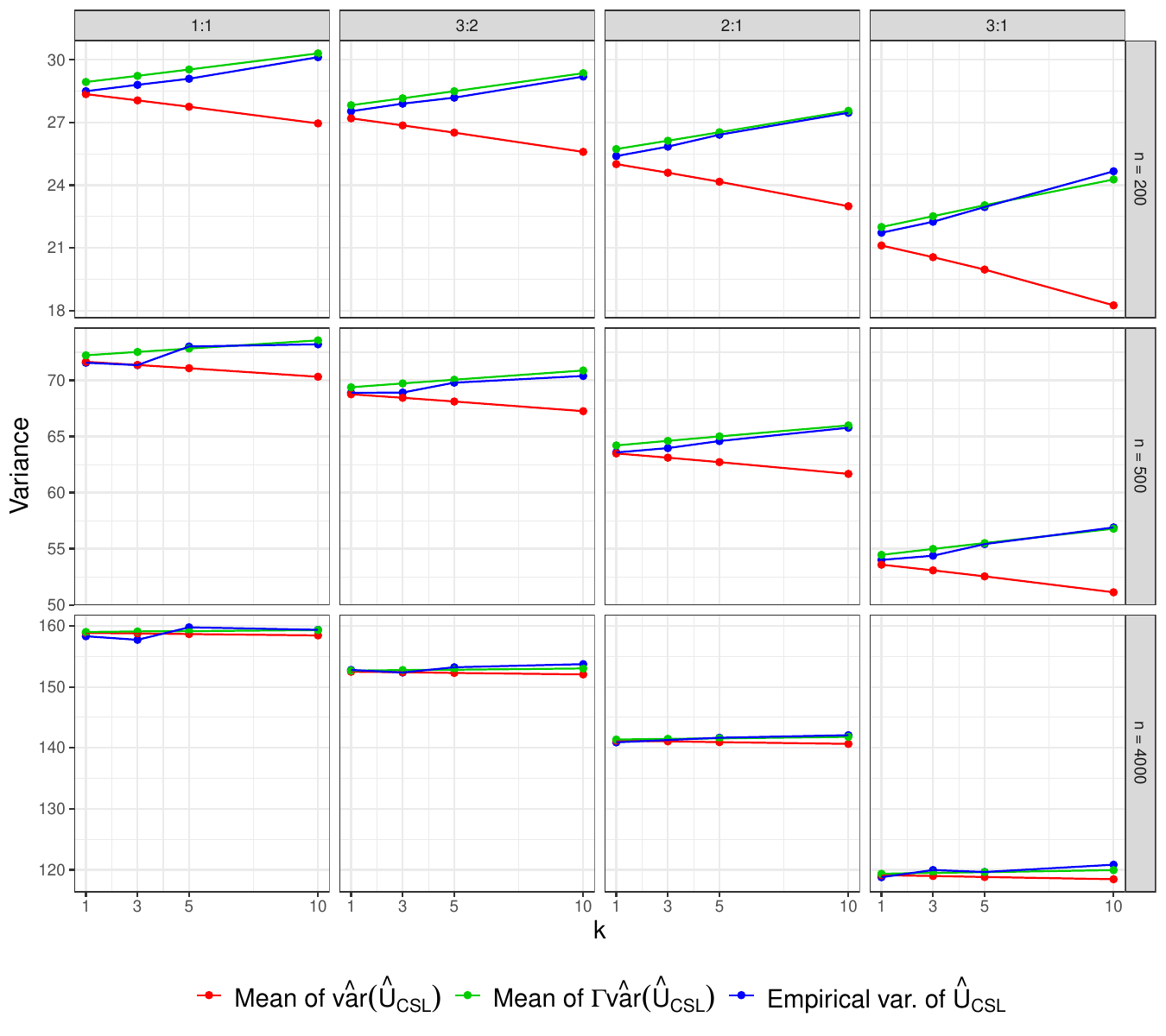}
    \caption{Empirical estimates of $\mathbb{E}\left\lbrace  \widehat{\operatorname{Var}}(\hat{U}_{CSL})\right\rbrace$  and  $\mathbb{E}\left\lbrace\Gamma\cdot \widehat{\operatorname{Var}}(\hat{U}_{CSL})\right\rbrace$, versus the empirical variance of $\hat{U}_{CSL}$ for varying number of adjustment covariates $k$. Results for different allocation ratios are shown in the columns, while different total sample sizes are presented in the rows. Based on 100.000 simulation runs.}
        \label{fig:results_var_strat}
\end{figure}

\begin{figure}[H]
    \centering
    \includegraphics[width=\linewidth]{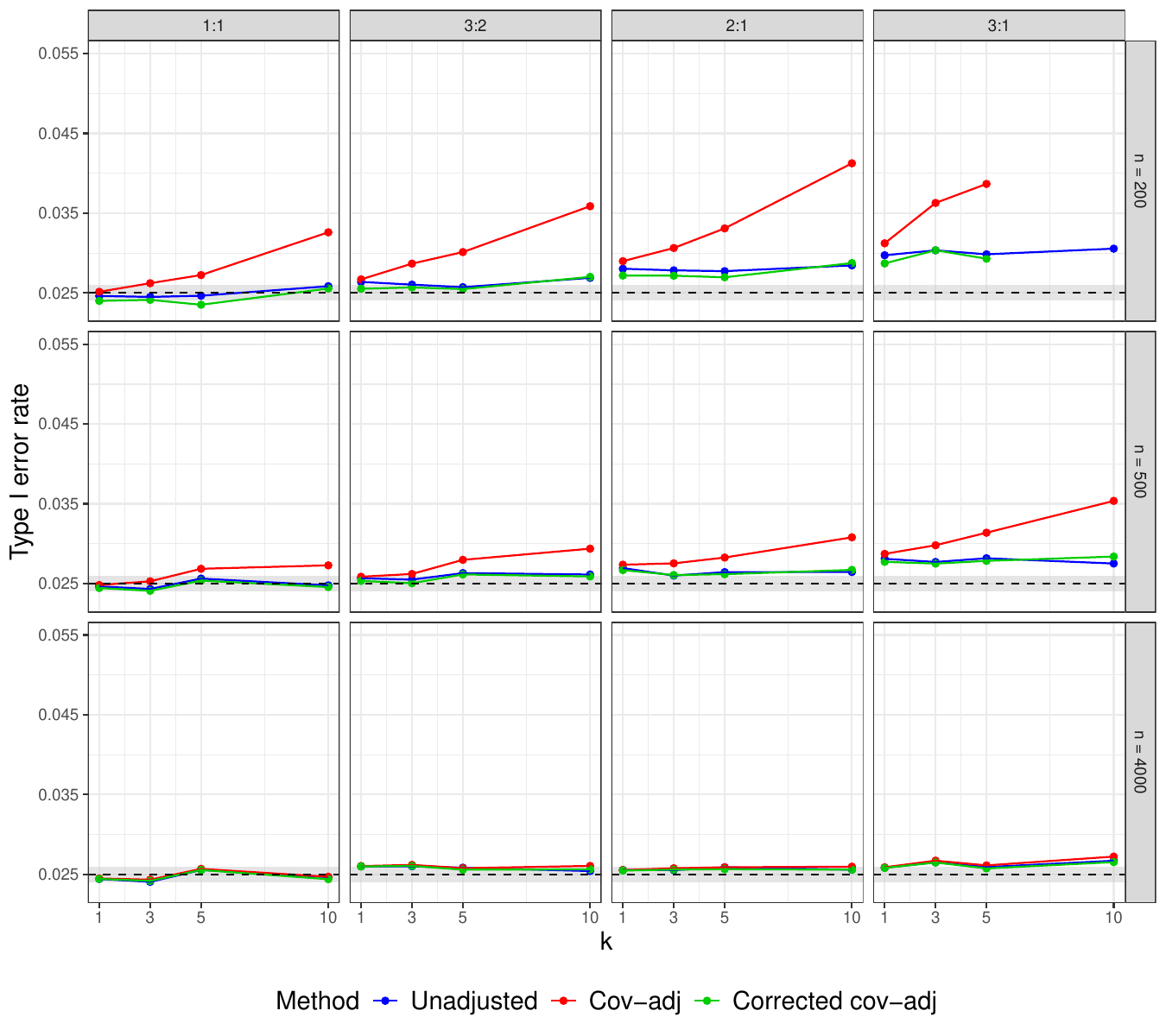}
    \caption{Type I error rate of the covariate-adjusted stratified log-rank test, the corrected covariate-adjusted stratified log-rank test, and the original stratified log-rank test (with no further covariate adjustment) for varying number of adjustment covariates $k$. Results for different allocation ratios are shown in the columns, while different total sample sizes are presented in the rows. Each plot includes a dashed reference line for the nominal significance level of 0.025, and the simulation error is shown as a gray area representing the 95\% confidence interval of the simulated type I error rate using 100.000 simulation runs. Type I error rate for the covariate-adjusted test with n = 200, 3:1 allocation, and k =  10 is not shown due to convergence issues.}
    \label{fig:results_t1e_strat}
\end{figure}

Ye et al. (\cite{ye2024covariate}) also propose a covariate-adjusted stratified log-rank test statistic
\begin{equation*}\label{eq_tcl}
    T_{CSL} = \hat{U}_{CSL}  / \sqrt{\widehat{\operatorname{Var}}}(\hat{U}_{CSL}),
\end{equation*}
where the covariate adjustment in $\hat{U}_{CSL}$ has the same structure as the covariate adjustment in $\hat{U}_{CL}$. Again, the same finite-sample correction term $\Gamma$ can be applied. In Figure \ref{fig:results_var_strat} we show the agreement between the empirical variance of $\hat{U}_{CSL}$ and the empirical mean of $\widehat{\operatorname{Var}}(\hat{U}_{CSL})$ with and without correction $\Gamma$ using the same scenarios as in Section \ref{sec:simulations} but with an additional covariate $Z_i$ entering the model as a stratification term, where $Z_i$ is generated independently from all other variables with a Bernoulli model with $P(Z_i = 1) = 0.3$. In Figure \ref{fig:results_t1e_strat} we show corresponding type I error rates for the tests.

\begin{figure}[H]
    \centering
    \includegraphics[width=\linewidth]{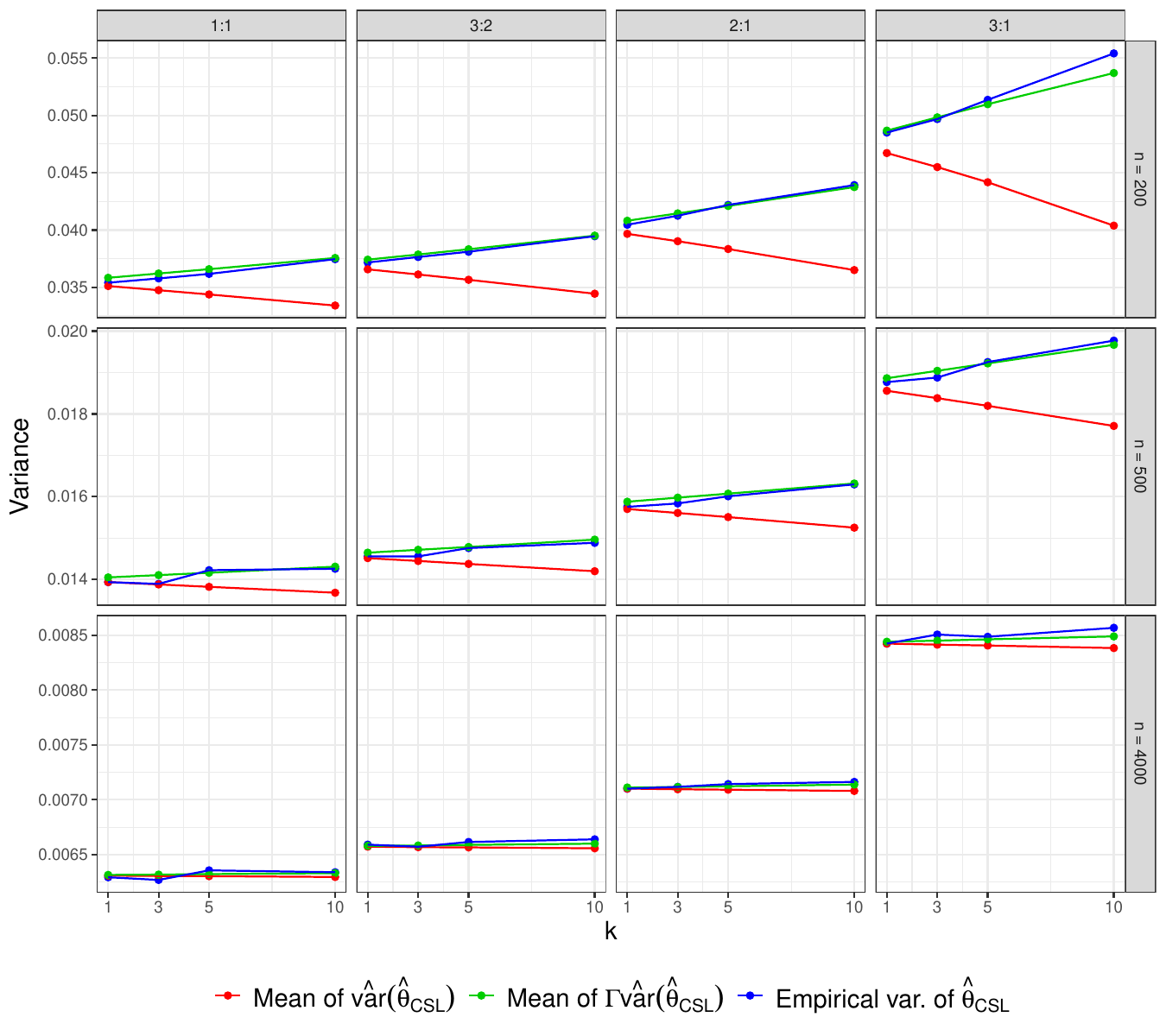}
    \caption{Empirical estimates of $\mathbb{E}\left\lbrace  \widehat{\operatorname{Var}}(\hat{\theta}_{CSL})\right\rbrace$  and  $\mathbb{E}\left\lbrace\Gamma\cdot \widehat{\operatorname{Var}}(\hat{\theta}_{CSL})\right\rbrace$, versus the empirical variance of $\hat{\theta}_{CSL}$ for varying number of adjustment covariates $k$. Results for different allocation ratios are shown in the columns, while different total sample sizes are presented in the rows. Based on 100.000 simulation runs.}
        \label{fig:results_var_lhr_strat}
\end{figure}

Finally, there exists a covariate-adjusted estimator $\hat\theta_{CSL}$ for the conditional log-hazard ratio (conditional on stratification variables), together with an associated variance estimator  $\widehat{\operatorname{Var}}(\hat{\theta}_{CSL})$. Once again, the same correction term $\Gamma$ can be applied. Figure \ref{fig:results_var_lhr_strat} shows the agreement between the empirical variance of $\hat{\theta}_{CSL}$ and the empirical mean of $\widehat{\operatorname{Var}}(\hat{\theta}_{CSL})$ with and without correction $\Gamma$ using the same scenarios as just described for the covariate-adjusted stratified log-rank test.

\section{Discussion}

While the covariate-adjusted log-rank test is an important contribution to statistical methodology for covariate adjustment, its finite-sample properties have not yet been systematically investigated. We demonstrate that the proposed estimator for the variance of the covariate-adjusted log-rank numerator tends to underestimate its true variance, leading to type I error rate inflation. This is particularly notable in trials with small sample sizes, unequal allocation, or a large number of adjustment covariates relative to the sample size. To mitigate this issue, we propose a finite-sample correction for the variance estimator, derived in the context of linear models with treatment-covariate interaction. The correction accounts for both the loss of residual degrees of freedom and the additional variance (i.e., the variance inflation factor) arising from the estimation of arm-specific covariate effects. Investigating scenarios with different sample sizes and allocation ratios, we show via simulations that applying the proposed correction substantially reduces the type I error rate inflation and generally leads to rejection rates comparable to those of an unadjusted log-rank test, where type I error rates are already known to be inflated under imbalanced allocation \cite{li2026issues}. 

\section*{Code}

The GitHub repository (\url{https://github.com/pavlakrotka/Cov-adj_log-rank_corr_paper}) contains the R code to reproduce the results of the simulation study.

\section*{Acknowledgements}

The motivation for the correction term $\Gamma(n_1, n_0, k)$ as well as its general structure is the work of the authors. LLMs assisted with deriving the concise numerical expressions for $\mathbb{E}(\mathrm{VIF})$ and $\mathbb{E}(Q_n)$ shown in the Appendix.

\bibliographystyle{ieeetr}
\bibliography{references}

\section*{Appendix}

\subsection*{Third term in variance estimator}

Let $Q_n
:=
(\widehat\beta_1-\widehat\beta_0)^\top
\widehat\Sigma_X
(\widehat\beta_1-\widehat\beta_0)$. 
Using the approximation $\widehat\Sigma_X\approx \Sigma_X$,
$$\mathbb E(Q_n)
\approx
\mathbb E\left(
(\widehat\beta_1-\widehat\beta_0)^\top
\Sigma_X
(\widehat\beta_1-\widehat\beta_0)
\right).$$
For any mean-zero random vector $Z$,
$$\mathbb E(Z^\top A Z)=\operatorname{tr}[A\operatorname{Var}(Z)].$$
Taking
$$Z=\widehat\beta_1-\widehat\beta_0,$$
and assuming $\beta_1=\beta_0$, we get
$$\mathbb E(Q_n)
\approx
\operatorname{tr}
\left[
\Sigma_X
\operatorname{Var}(\widehat\beta_1-\widehat\beta_0)
\right].$$
If the two arm-specific slope estimators are asymptotically independent, then
$$\operatorname{Var}(\widehat\beta_1-\widehat\beta_0)
\approx
\operatorname{Var}(\widehat\beta_1)
+
\operatorname{Var}(\widehat\beta_0).$$
Therefore,
$$\mathbb E(Q_n)
\approx
\operatorname{tr}
\left[
\Sigma_X
\left(
\operatorname{Var}(\widehat\beta_1)
+
\operatorname{Var}(\widehat\beta_0)
\right)
\right].$$
If each arm uses ordinary least squares with $k$ covariates and homoskedastic residual variance $\sigma^2$, then approximately
$$\operatorname{Var}(\widehat\beta_1)
\approx
\frac{\sigma^2}{n_1}\Sigma_X^{-1},$$
and
$$\operatorname{Var}(\widehat\beta_0)
\approx
\frac{\sigma^2}{n_0}\Sigma_X^{-1}.$$
Plugging these in,
$$\mathbb E(Q_n)
\approx
\operatorname{tr}
\left[
\Sigma_X
\left(
\frac{\sigma^2}{n_1}\Sigma_X^{-1}
+
\frac{\sigma^2}{n_0}\Sigma_X^{-1}
\right)
\right].$$
Since
$$\operatorname{tr}(\Sigma_X\Sigma_X^{-1})=k,$$
we obtain
$$\mathbb E(Q_n)
\approx
\sigma^2 k
\left(
\frac{1}{n_1}
+
\frac{1}{n_0}
\right).$$

\subsection*{Expectation of the variance inflation factor}

The variance inflation factor is given by:

$$\mathrm{VIF} = \frac{c^\top (\mathbf{X}^\top \mathbf{X})^{-1} c}{1/n_1 + 1/n_0},$$

where $\mathbf{X}$ is the model matrix, and $c = (1, -1, \bar{X}, - \bar{X})^\top$ is the contrast vector.

The model matrix $\mathbf{X}$ can be written as

$$\mathbf{X}
=
\begin{pmatrix}
A_i & 1-A_i & A_i X_i^\top & (1-A_i)X_i^\top
\end{pmatrix}.$$

If we order the observations by treatment arm, $\mathbf{X}$ becomes block diagonal, such that

$$\mathbf{X}^\top \mathbf{X}
=
\begin{pmatrix}
Z_1^\top Z_1 & 0 \\
0 & Z_0^\top Z_0
\end{pmatrix},$$

where

$$Z_1 =
\begin{pmatrix}
1 & X_i^\top
\end{pmatrix}_{A_i=1}
\qquad
Z_0 =
\begin{pmatrix}
1 & X_i^\top
\end{pmatrix}_{A_i=0}$$

Therefore,

$$c^\top (\mathbf{X}^\top \mathbf{X})^{-1} c
=
d^\top (Z_1^\top Z_1)^{-1} d
+
d^\top (Z_0^\top Z_0)^{-1} d,$$

where $d =
\begin{pmatrix}
1 \\
\bar X
\end{pmatrix}.$

Using the standard leverage identity for linear regression with intercept gives:

$$d^\top (Z_j^\top Z_j)^{-1} d
=
\frac{1}{n_j}
+
(\bar X - \bar X_j)^\top S_j^{-1}(\bar X - \bar X_j),$$

where $S_j
=
\sum_{i:A_i=j}
(X_i - \bar X_j)(X_i - \bar X_j)^\top.$

Note that 
$\bar X
=
(n_1 \bar X_1 + n_0 \bar X_0) / n,$
and hence
$\bar X - \bar X_1
=
\frac{n_0}{n}(\bar X_0 - \bar X_1).$ Therefore,

$$d^\top (Z_1^\top Z_1)^{-1} d
=
\frac{1}{n_1}
+
\left(\frac{n_0}{n}\right)^2
(\bar X_0 - \bar X_1)^\top
S_1^{-1}
(\bar X_0 - \bar X_1).$$

Similarly, for the control group,

$$d^\top (Z_0^\top Z_0)^{-1} d
=
\frac{1}{n_0}
+
\left(\frac{n_1}{n}\right)^2
(\bar X_1 - \bar X_0)^\top
S_0^{-1}
(\bar X_1 - \bar X_0).$$

Therefore,

$$c^\top (\mathbf{X}^\top \mathbf{X})^{-1} c
=
\frac{1}{n_1}
+
\frac{1}{n_0}
+
\left(\frac{n_0}{n}\right)^2
\Delta^\top S_1^{-1}\Delta
+
\left(\frac{n_1}{n}\right)^2
\Delta^\top S_0^{-1}\Delta,$$

where
$\Delta = \bar X_1 - \bar X_0.$

This form enables us to compute the expectation $\mathbb E(c^\top (\mathbf{X}^\top \mathbf{X})^{-1} c)$. Since $X_i \sim N(0, I_k)$, $\Delta \sim N\left(0, \left(\frac{1}{n_1}+\frac{1}{n_0}\right)I_k\right).$
Moreover, the matrix $S_j$ follows the Wishart distribution with scale matrix $I_k$, and $n_j-1$ degrees of freedom:
$S_1 \sim W_k(I_k, n_1-1)$
and
$S_0 \sim W_k(I_k, n_0-1)$.

Using the expectation of the inverse Wishart distribution, it follows that
$$\mathbb E(S_j^{-1})
=
\frac{I_k}{n_j - k - 2},$$

provided $n_j > k + 2$.

Moreover, note that
$\mathbb E\left(\Delta^\top S_1^{-1}\Delta\right)
=
\operatorname{tr}
\left[
\mathbb E(S_1^{-1}) \mathbb E(\Delta\Delta^\top)
\right].$

Since
$\mathbb E(\Delta\Delta^\top)
=
\left(\frac{1}{n_1}+\frac{1}{n_0}\right)I_k,$
it follows that
$\mathbb E\left(\Delta^\top S_1^{-1}\Delta\right)
=
\frac{k}{n_1 - k - 2}
\left(\frac{1}{n_1}+\frac{1}{n_0}\right)$
and
$\mathbb E\left(\Delta^\top S_0^{-1}\Delta\right)
=
\frac{k}{n_0 - k - 2}
\left(\frac{1}{n_1}+\frac{1}{n_0}\right).$

Hence, the expectation of $c^\top (\mathbf{X}^\top \mathbf{X})^{-1} c$ is given by

\begin{flalign*}
\mathbb E\left(
c^\top (\mathbf{X}^\top \mathbf{X})^{-1} c
\right)
& =
\frac{1}{n_1}
+
\frac{1}{n_0}
+
\left(\frac{n_0}{n}\right)^2
\frac{k}{n_1 - k - 2}
\left(\frac{1}{n_1}+\frac{1}{n_0}\right)
+
\left(\frac{n_1}{n}\right)^2
\frac{k}{n_0 - k - 2}
\left(\frac{1}{n_1}+\frac{1}{n_0}\right) \\
& = \left(\frac{1}{n_1}+\frac{1}{n_0}\right)
\left(
1
+
k
\left(
\frac{(n_0/n)^2}{n_1-k-2}
+
\frac{(n_1/n)^2}{n_0-k-2}
\right)
\right)
\end{flalign*}

Dividing by the unadjusted variance factor $\frac{1}{n_1}+\frac{1}{n_0}$ leads to

$$\mathbb E(\mathrm{VIF})
=
1
+
k
\left(
\frac{(n_0/n)^2}{n_1 - k - 2}
+
\frac{(n_1/n)^2}{n_0 - k - 2}
\right).$$

Moreover, note that for balanced allocation, this reduces to

$$\mathbb E(\mathrm{VIF})
=
1
+
k
\left(
\frac{1/4}{m - k - 2}
+
\frac{1/4}{m - k - 2}
\right)
=
1
+
\frac{k}{2(m -k-2)},
$$

where $n_0 = n_1 = m$.

\end{document}